\documentclass[12pt]{JHEP3}
\usepackage{amssymb}
\usepackage{amsmath}
\usepackage{epsfig}

\title{ Holograph in noncommutative geometry: Part 1}
\author{Jingbo Wang \\
  Department of Applied Physics,
 Xi'an Jiaotong University,
 Xi'an, 710049, China\\Kavli Institute for Theoretical Physics China, CAS, Beijing
  100190, China
 \email{ shuijing31@gmail.com }}
 \abstract { In this paper, we consider the holograph principle
 emergent from noncommutative geometry, based on the spectral action principle.
 We show that under some appropriate conditions, the gravity theory on a manifold with boundary could be equivalent to
 a gauge theory $SU(N)$ on the boundary. Then an expression for $N$ with the geometrical quantities of the
 manifold is given. Based on this result, we find that the volume
 of the manifold and the boundary have some discrete structure.
 Applying the result to the black hole, we get that the radium of
 the Schwarzschild black hole is quantized. We also find an
 explanation why the extremal RN-black hole has zero temperature but
 with finite entropy.}
 \preprint{} \dedicated{}
\begin{document}
\section{Introduction}
Noncommutative geometry(NCG)\cite{ar1} is a branch of mathematics
that has many applications in physics, such as in string
theory\cite{ar2}, quantum Hall effect\cite{ar3}, cosmology
\cite{ar4}and so on.It can unify the gravity theory and the gauge
theory, and also can afford a starting point for quantum gravity. A
good review of the noncommutative space appeared in physics is
\cite{ar5}.

It is well believed that the holograph principle \cite{ar6,ar7}is a
fundamental principle in quantum gravity. It roughly states that
quantum gravity on a bulk space can be described by a suit theory on
the boundary. This principle is motivated by the black hole physics,
and strengthened by the AdS/CFT duality in M theory.

Since the NCG is an approach to quantum gravity, we can conjecture
that the holograph principle will emergent from it as well. In this
paper we will show that the holograph principle is a result of the
spectral action principle on suit system. In this paper, we work
with Planck unit system $\hbar =c=1$.
\section{Two systems}
Let $M$ be a ($n+1$) dimension smooth compact Riemannian manifold
with smooth boundary $\partial M$. Then $\partial M$ has dimension
$n$. Assume $n$ to be even. Let $\Delta$ be the scale Laplace
operator on $M$. Define the eigenvalue problem:
\begin{eqnarray}
\Delta \phi_n & =& \lambda_n \phi_n,\label{1}\\
\mathcal {B}^\pm \phi &=& 0. \end{eqnarray} where $\mathcal {B}$ is
called the boundary operator. The frequently used condition are the
Dirichlet boundary condition $\mathcal {B}^-$ and Neumann boundary
condition $\mathcal {B}^{+}$ respectively:
\begin{eqnarray}
\mathcal {B}^- \phi &=& \phi |_ {\partial M}\label{2}\\
\mathcal {B}^{+} \phi &=& (\phi_{;n}+S\phi) |_{\partial M}\label{3},
\end{eqnarray}
where S is a matrix valued function on $\partial M$. The boundary
condition \ref{3} is also called Robin condition.

For the scale Laplace operator $\Delta$ on the manifold $M$, we
define its spectral action:
\begin{eqnarray}
I=Tr e^{-\Delta/\Lambda^2}\label{4}. \end{eqnarray} Then it has a
heat kernel expansion \cite{ar8}:
\begin{eqnarray}
    I=Tr e^{-\Delta/\Lambda^2}\simeq \sum \limits_{m\geqslant 0}
    {\Lambda^{(n+1)-m}}a_m(\Delta).
\end{eqnarray}

where the Seeley-de Witt coefficients $a_m$ are given by \cite{ar8}:
\begin{eqnarray}
    a_0(\Delta,\mathcal {B})=(4\pi)^{-(n+1)/2}\int_M{d^{(n+1)}x\sqrt
    g}=(4\pi)^{-(n+1)/2}vol(M).\label{5}\\
a_1(\Delta,\mathcal {B}) = 1/4\chi(4\pi)^{-n/2}\int_{\partial
M}{d^nx\sqrt h }= 1/4\chi(4\pi)^{-n/2}vol(\partial M).
\end{eqnarray} where $h$ is the induced metric on the boundary, and
$\chi=\pm1$ for boundary condition $\mathcal {B^{\pm}}$. This
spectral action contain the cosmology term, the Hilbert-Einstein
term, and other high order terms.

 Then we consider another
system, the gauge theory $SU(N)$ on the boundary $\partial M$,
considered in paper\cite{ar9}. The spectral triple for this system
is $(\mathcal {A},\mathcal {H},D)$,where
\begin{eqnarray}
    \mathcal {A}=C^\infty(\partial M)\otimes M_N(C),
    \mathcal {H}=L^2(\partial M,S)\otimes M_N(C),
    D=\partial _{\partial M} \otimes 1.
\end{eqnarray} After inner fluctuation the Dirac operator is given by
\begin{eqnarray}
    D=e^\mu_a \gamma^a((\partial _\mu+\omega_\mu)\otimes 1_N+1\otimes (-\frac{\imath}{2} g_0 A^i_\mu
    T^i)).
\end{eqnarray} where $\omega_\mu$ is the spin connection on $\partial
M$, and $T^i$ are matrices in the adjoint representation of $SU(N)$
satisfying $Tr(T^i T^j)=2\delta^{ij}.$

Then we can calculate the spectral action for this system, and the
result was given in paper \cite{ar9}:
\begin{equation}\label{7}
    I'=Tr e^{-D^2/\Lambda^2}\simeq \sum \limits_{m\geqslant 0}
    {\Lambda^{m-n}}\int_{\partial M}{a_m(x,D^2)dv(x)}.
\end{equation}
The coefficients $a'_m(D^2)$ are:
\begin{eqnarray}
a'_0(x,D^2)&=& (4\pi)^{-n/2} Tr(I),\nonumber\\
a'_2(x,D^2)&=& (4\pi)^{-n/2} Tr(R/6I+E),\nonumber\\
a'_{2m+1}(x,D^2)&=&0 .\label{8} \end{eqnarray} So we can get:
\begin{eqnarray}
a'_0(D^2)&=&(4\pi)^{-n/2} N 2^{n/2}\int_{\partial
M}{d^nx\sqrt h},\nonumber\\
a'_2(x,D^2)&=&-\frac{N 2^{n/2} }{12(4\pi)^{n/2}}  \int_{\partial
M}{d^nx\sqrt h R},\nonumber\\
a'_{2m+1}(x,D^2)&=& 0 .\label{81}\end{eqnarray}

Now we have two systems: on the one hand, the scale Laplace operator
on the manifold $M$ with boundary $\partial M$ which describes pure
gravity; on the other hand, the square of the Dirac operator which
contain the gauge part on the boundary $\partial M$. According to
the spectral action\cite{ar9}, that is "the physical action only
depends upon the spectrum", if those two system have the same
spectrum, hence the same spectral action, they describe the same
physics. From this we get can the holograph principle. So next we
will investigate under what conditions we can get the same spectral
action.
\section{The Correspondence}
In this section, we will see under what conditions the two spectral
action coincide. First let's introduce a length parameter $l_c\equiv
\Lambda^{-1}$. If the two spectral action coincide, we have:
\begin{eqnarray}
a_0 l_c^{-(n+1)}+a_1 l_c^{-n}&=& a'_0 l_c^{-n},\nonumber\\
a_2 l_c^{-(n-1)}+a_3 l_c^{-n+2}&=& a'_2 l_c^{-n-2},\nonumber\\
a_n l_c^{-1}+a_{n+1}&=& a'_n ,\label{10}
\end{eqnarray} From the first equation, we get:
\begin{equation}\label{111}
    vol(M)=vol(\partial M) \sqrt{4\pi}(N 2^{n/2}-1/4\chi) l_c.
\end{equation}
That is  \begin{equation}\label{11}
    N=\frac{vol(M)}{vol(\partial M) \sqrt{\pi} 2^{n/2+1}l_c}+2^{-n/2-2}\chi.
\end{equation}Since $N$ is a positive integer, the equation
indicate that the volume of the manifold and the boundary must have
some discrete structure.

Remark 1: We can see that the expression for the $N$ has two terms,
one term containing $l_c$, and the other not. This splitting is
essential for the holograph. In $I'$, the odd term $a_{2n+1}$ is
zero, and $N$ will split into two terms to correspond to the bulk
and boundary terms in $I$. This is the reason why we use the scale
Laplace operator instead of the more nature Dirac operator. For
Dirac operator, the $a_1=0$ due to the $Trace(\chi)=0$\footnote{The
author will thank Alain Connes for explanation on this point.}, but
other odd terms are not zero. Due to \ref{11}, the $N$ only has one
term, and can't get the odd terms in $I$.

This is our first result from our assumptions, and we want to see
the consequence of it.
\section{Black hole entropy}
From black hole thermodynamics we know that a black hole have a
temperature and an entropy \cite{ar10}. First let's consider the
simplest one, the Schwarzschild black hole:
\begin{eqnarray}\label{12}
    ds^2 = -(1-2M/r)dt^2+(1-2M/r)^{-1}dr^2 +r^2(d\theta^2+sin^2\theta
    d\varphi^2).
\end{eqnarray}
which has a temperature $T_{BH}=\frac{1}{4\pi r_s}=\frac{1}{8\pi
M}$, and entropy $S=\frac{A}{4l_p^2}$.

Consider the black hole as the region inside the horizon and the
horizon as the boundary, we apply the above result. In this case,
$n=2$.

The volume inside the horizon is
\begin{equation}\label{13}
    vol(M)=\int_M{d^3x\sqrt {|g|}}=4\pi
    \int_0^{r_s}{\frac{r^2dr}{\sqrt{r_s/r-1}}}.
\end{equation}
So we can get
\begin{equation}\label{131}
    vol(M)=5/4\pi^2r_s^3.
\end{equation}
 With induced metric on the boundary, we can
get $vol(\partial M)=4\pi r_s^2$. Take those values into equation
\ref{111}, we can get
\begin{equation}\label{14}
    5/4\pi^2 r_s^3=4\pi r_s^2 \sqrt{4\pi}(2N-1/4\chi)l_c.
\end{equation}
Then we get
\begin{equation}\label{15}
    r_s=32/5\pi^{-1/2}(2N-1/4\chi) l_c.
\end{equation}
So the smallest radius of a black hole is
$(r_s)_{min}=32/5\pi^{-1/2}(2-1/4\chi) l_c$ for different boundary
condition. And the gap between nearest radius is
\begin{equation}\label{151}
    \Delta r=64/5 \pi^{-1/2} l_c.
\end{equation}
We can see that the radium is equidistant, so the area is not. This
result is sharply contrary to Bekenstein's suggestion\cite{ar11}:
$A=\alpha n l_p^2.$

 Change the expression \ref{15} to another way, we get
\begin{equation}\label{16}
    N=\frac{5 \sqrt \pi r_s}{64 l_c}+\chi/8.
\end{equation}

Consider the gauge theory on the horizon. We have an expression for
$N$ with the geometry of the black hole, the equation \ref{16}. With
some basic quantum statistics calculations, we can get the entropy
for the free $SU(N)$ glues gases in two dimension at temperature
$T$, that is
\begin{equation}\label{17}
    S=\frac{3AT^2}{4\pi}(N^2-1)\zeta(3).
\end{equation}
With the expression for $T=T_{BH}$ and $N$, we get
\begin{equation}\label{18}
    S=\frac{3\zeta(3)}{64^2\pi^2 }(\frac{25A}{64l_c^2}+2.5\sqrt A/l_c-252)
    \qquad   N\geqslant 2.
\end{equation}
This is the entropy of free glues gas, for interacting glues gas, we
have to add a factor $S'=f(g)S$, where $f(g)$ is a function
depending on the coupling constant. The gauge coupling constant $g$
will appear in $a_4$ term, so we don't know which limit of gravity
theory this weak limit of gauge theory correspondence to. We know
that the entropy for a black hole is expression
$S=\frac{A}{4l_p^2}$, so they have the same structure, but we don't
know the exact relation between $l_c$ and $l_p$.

From equation \ref{18}, we can find that the entropy has a $\sqrt A$
term, which is absent in the usual formula. But since there maybe
exist the volume correction for the black hole entropy \cite{ar12},
the length correction maybe also exist.

Since there exist the smallest length of black hole, we can
calculate the smallest area for a Schwarzschild black hole.
\begin{eqnarray}
    A_{min}=4\pi (r_s)_{min}^2=(112/5l_c)^2.
    \end{eqnarray}
 We assume
that the $l_c$ is a fundamental length scale in noncommutative
geometry.

Next we consider the Reissner-Nordstrom(RN) black hole, which has
the metric:
\begin{eqnarray}
  ds^2 = -(1-2M/r+Q^2/r^2)dt^2+(1-2M/r+Q^2/r^2)^{-1}dr^2
    +r^2(d\theta^2+sin^2\theta
    d\varphi^2).
\end{eqnarray}
There are two horizons with radium
\begin{equation}\label{20}
    r_{\pm}=M\pm \sqrt{M^2-Q^2}.
\end{equation}
The extremal RN-black hole has $r_+=r_-$. The temperature on the
horizons are:
\begin{equation}\label{21}
    T_{\pm}=\frac{r_+ - r_-}{4\pi r_{\pm}^2}.
\end{equation}
The entropy of the RN-black hole is $S=\frac{A_+^2}{4l_p^2}.$

We apply our method to the RN-black hole. Consider the black hole as
the region between the two horizons, and the horizons as the
boundary with opposite direction. The volume between the two
horizons is
\begin{equation}\label{22}
    vol(M)=\int_M{d^3x\sqrt {|g|}}=4\pi
    \int_{r_-}^{r_+}{\frac{r^3dr}{\sqrt{(r-r_-)(r_+ -r)}}}.
\end{equation}
Redefine $r_-=a,\qquad r_+ -r_- =b$. We can calculate this integral
to get
\begin{equation}\label{23}
    vol(M)=4\pi^2(a^3+3/2a^2b+9/8ab^2+5/16b^3).
\end{equation}
When $r_-=a=0$, then $r_+=b=2M$, that is the Schwarzschild black
hole, we get $vol(M)=5/4\pi^2 r_+^3$, just the result \ref{131}.
Since the boundary have opposite direction, the boundary should be
\begin{eqnarray}\label{24}
    vol(\partial M)=4\pi(A_+ -A_-)=4\pi(r_+^2-r_-^2)
    =4\pi (2a+b) b.
\end{eqnarray}
Putting those expression into equation \ref{11}, we can get
\begin{equation}\label{25}
    N=\frac{(a^3+3/2a^2b+9/8ab^2+5/16b^3)\sqrt{\pi}}{4(2a+b) b
    l_c}+\chi/8
\end{equation}

Let's consider the extremal RN-black, that is $b\rightarrow 0$, then
$T\rightarrow 0$. On the other hand,
\begin{equation}\label{26}
    N\approx \frac{(a^3+3/2a^2b)\sqrt{\pi}}{4(2a+b) b
    l_c}\rightarrow \infty.
\end{equation}
The entropy of the $SU(N)$ glues gas on the two boundary is
\begin{eqnarray}
S = \frac{3\zeta(3)}{4\pi}(A_+ T_+^2 +A_- T_-^2)N^2
\approx\frac{3\zeta(3)}{4^5\pi^3}(Nb/(a+b))^2\approx\frac{3\zeta(3)A_-^2}{2^{11}\pi^2l_c^2}.\label{27}
\end{eqnarray}
From the above calculation we can see that in the extremal case,
though the temperature approach zero, on the other hand, the $N$
approach infinite, and their product is a finite number, so the
entropy is finite too. To describe an extremal RN-black hole, we
have to use the $SU(\infty)$ gauge theory. Since the entropy of the
glues gas is related to the black hole entropy, the black hole
entropy is also finite.

For a general RN-black hole, we can also calculate the entropy. At
this time,\begin{equation}\label{28}
    N\approx \frac{(5r_+^2-2r_+ r_- +5r_-^2)\sqrt{\pi}}{64(r_+-r_-)
    l_c}.
\end{equation}
So the entropy is
\begin{eqnarray}
S = \frac{3\zeta(3)}{4\pi}(A_+ T_+^2 +A_- T_-^2)N^2
\approx\frac{3\zeta(3)A_+}{2^{17}\pi^2l_c^2}(5-2r_-/r_++5r_-^2/r_+^2)^2.\label{29}
\end{eqnarray}
So in the general case, the entropy is a function of both $r_+$ and
$r_-$.
\section{Conclusion}
In this paper, we consider two related systems: the gravity theory
on a manifold with boundary, and the gauge theory on the boundary.
Under some conditions, the two system may have the same spectral
action. Due to the spectral action principle, they have the same
physics, so we get the holograph principle.

In this paper, we only consider the first term in the spectral
action expansion, and get the relation between the geometrical
quantities on the gravity side and $N$ on the gauge side. We then
apply this relation to black hole entropy, and get a qualitative
right expression.

In the following work, we will calculate the next few terms in the
spectral action expression, and get more relations between the
gravity theory and the gauge theory.

We think that the spectral action principle is a more fundamental
principle in quantum gravity, and may play some essential role in
the background-independent formulation of M theory. We should pay
much more attention to it.
\begin{acknowledgments}
The author will thank Doctor Zhiwen Shi and Doctor Pong Song in
institute of physics (Chinese Academy of Science) for many help.
\end{acknowledgments}


\begin{thebibliography}{99}
\bibitem{ar1} A.Connes, {\it Noncommutative Geometry,} Academic
Press, New York, 1994
\bibitem{ar2} N.Seiberg and E.Witten,\jhep{09}{1999}{032}
\bibitem{ar3} J.Bellissard, A.van Elst, and H.Schulz-Baldes,
\jmp{35}{1994}{5373-5451}
\bibitem{ar4} M.Marcolli and E.Pierpaoli, {\it Early universe models
from noncommutative geometry}, \arXivid{0908.3683}
\bibitem{ar5} A.Connes and M.Marcolli, {\it A walk in the
noncommutative garden}, \Math{QA}{0601054}
\bibitem{ar6} G.'t Hooft, {\it Dimensional reduction in quantum
gravity}, \grqc{9310026}
\bibitem{ar7} L.Susskind, \jmp{36}{1995}{6377}
\bibitem{ar8} D.Vassilevich, \prep{388}{2003}{279}
\bibitem{ar9} Ali H.Chamseddine, A.Connes, \cmp{186}{1997}{731}
\bibitem{ar10} S.Carlip, {\it Black hole thermodynamics and
statistical mechanics}, \arXivid{0807.4520}
\bibitem{ar11} J.D.Bekenstein {\it Black holes: classical properities,
thermodynamics, and heuristic quantization}, \grqc{9808028}
\bibitem{ar12} E.R.Livine, D.R.Terno, \npb{794}{2008}{138}
\end{thebibliography}
\end{document}